\documentstyle[11pt,epsf]{article}
\newtheorem{theorem}{Theorem}

\newcommand {\dfn} {\stackrel{\Delta} {=}}

\newcommand {\reals} {{\rm I\!R}}

\newcommand {\bu} {\mbox{\boldmath $u$}}
\newcommand {\bv} {\mbox{\boldmath $v$}}

\newcommand {\bx} {\mbox{\boldmath $x$}}
\newcommand {\by} {\mbox{\boldmath $y$}}

\newcommand {\bE} {\mbox{\boldmath $E$}}

\newcommand{\calI}{{\cal I}}

\newcommand{\calU}{{\cal U}}
\newcommand{\calV}{{\cal V}}

\newcommand{\calX}{{\cal X}}
\newcommand{\calY}{{\cal Y}}
\newcommand{\calZ}{{\cal Z}}

\begin{document}
\thispagestyle{empty}
\title{On the Data Processing Theorem in the Semi--Deterministic Setting}
\author{Neri Merhav}
\date{}
\maketitle

\begin{center}
Department of Electrical Engineering \\
Technion - Israel Institute of Technology \\
Technion City, Haifa 32000, ISRAEL \\
E--mail: {\tt merhav@ee.technion.ac.il}\\
\end{center}
\vspace{1.5\baselineskip}
\setlength{\baselineskip}{1.5\baselineskip}

\begin{abstract}

Data processing lower bounds on the expected distortion are derived in the
finite--alphabet semi--deterministic setting, where the source produces a deterministic,
individual sequence, but the channel model is probabilistic, and the decoder
is subjected to various kinds of limitations, e.g., decoders implementable by
finite--state machines, with or without counters, and with or without
a restriction of common reconstruction with high probability. Some of our
bounds are given in terms of the Lempel--Ziv complexity of the source sequence
or the reproduction sequence. We also
demonstrate how some analogous results can be obtained for
classes of linear encoders and linear decoders in the continuous alphabet case.\\

\vspace{0.3cm}

\end{abstract}

\section{Introduction}

In a series of articles from the
seventies and the eighties of the twentieth century, Ziv
\cite{Ziv78},\cite{Ziv80},\cite{Ziv84}, and Ziv and Lempel
\cite{LZ86}, \cite{ZL78},
have created a theory of universal source coding for
individual sequences using finite--state machines.
In particular, the work \cite{Ziv78}
focuses on universal, fixed--rate, (almost)
lossless compression of individual sequences
using finite--state encoders and decoders,
which was then further developed to the
famous Lempel--Ziv algorithm \cite{LZ86}, \cite{ZL78}.
In \cite{Ziv80}, the framework of \cite{Ziv78} was extended to
lossy coding for both noiseless and noisy transmission (subsections II.A and
II.B of \cite{Ziv80}, respectively), and later
further extended in other directions, such as incorporation of side information
in the context of almost lossless compression, where the side information data
is also modeled as an individual sequence \cite{Ziv84}, in other words, an
individual--sequence counterpart of Slepian--Wolf coding \cite{SW73}
was studied in \cite{Ziv84} (see also a later extension to the lossy case \cite{MZ06}).

The main trigger for this paper stems from the coding
theorem for noisy 
transmission in \cite[Subsection II.B]{Ziv80}. We begin by revisiting the
assertion and the proof of the converse part of this theorem (Theorem 3 and
eqs.\ (12) and (13) in \cite{Ziv80}), which provides a lower bound on the
distortion in a {\it
semi--deterministic setting}, where the source emits a deterministic
(individual) sequence, but the channel model is probabilistic as usual (in
particular, it is a discrete memoryless channel) and the encoder and decoder are
limited to be finite--state machines with no more than $s$ states and a given
overall delay, which we shall denote by $d$. While this theorem is essentially
correct, it turns out that there are certain imprecise steps in its proof
(see Appendix for details)
and moreover, in relation to our corrections to this proof, the assertion of
the theorem itself can be strengthened and sharpened. 
The revisited converse theorem
imposes no limitations on the encoder,\footnote{The assumption that the encoder
is a finite--state machine is not really used in \cite{Ziv80} either,}
and allows the decoder to be equipped
with a modulo--$\ell$ counter ($\ell$ -- positive integer) in addition to its
$s$ states of memory, which means that within each period of length $\ell$,
the decoder is allowed to be time--varying, as opposed to the time--invariant
model used in \cite{Ziv80} and in related papers.\footnote{One might argue that a
finite--state machine with $s$ states and a modulo--$\ell$ counter is just
a particular finite--state machine with a total number of $s\cdot\ell$ states.
While this argument is
true, in principle, the idea is that this partition of the total number of allowed states between
those that are allocated to implement a clock (the counter) and those that are allocated
to memory of past input data (the remaining $s$ states) give us more detailed and more refined
results.} Also,
our lower bound on the distortion depends, not
only on the number of states $s$ (as in \cite{Ziv80}), but also on the allowed
delay $d$ (as well as on some additional redundancy terms).

Beyond the above described revisit of Theorem 3 of \cite{Ziv80}, we also
derive additional lower bounds on the expected distortion in the
semi--deterministic setting. One of them
is associated with a restriction of a {\it common reconstruction} (with 
high probability) at both encoder and decoder, which is a setup that has recently received some attention
in other contexts, like the Wyner--Ziv problem (see e.g., \cite{Steinberg09}), with motivations
in medical imaging, etc. In addition, some of our
bounds are given in a
more explicit form, in terms of the Lempel--Ziv complexity of the source sequence
or the reproduction sequence. This may be interesting in the sense that the
Lempel--Ziv complexity usually arises when the finite--state structure is
imposed on the encoder, whereas in our case, it is imposed on the decoder.
Finally, we
demonstrate how some analogous results can be obtained for
classes of linear encoders and linear decoders in the continuous alphabet
case. 

It should be emphasized that our focus in this paper is primarily on
lower bounds and converse theorems, and not quite on achievability schemes.
Most of our bounds can 
be asymptotically approached by conceptually simple, separation--based
schemes, in the spirit of the one proposed in \cite{Ziv80} or with certain
modifications and 
variations on the same ideas.

The outline of this paper is as follows. In Section 2, we establish notation
conventions and formalize the semi--deterministic setting under consideration.
In Section 3, we derive a lower bound on the distortion without the common reconstruction
requirement, and in Section 4, we derive the parallel lower bound under common reconstruction.
In both sections, we also derive the aforementioned alternative lower bounds, which can be
calculated more easily.
Finally, in Section 5, we give an outline of an analogue of the main result of
Section 2 for continuous alphabets and linear encoders and decoders.

\section{Problem Formulation and Notation Conventions}

Throughout the paper, random variables will be denoted by capital
letters, specific values they may take will be denoted by the
corresponding lower case letters, and their alphabets
will be denoted by calligraphic letters. Similarly, random
vectors, their realizations, and their alphabets, will be denoted,
respectively, by capital letters, the corresponding lower case letters,
and calligraphic letters, all superscripted by their dimensions. For
example, the random vector $Y^n=(Y_1,\ldots,Y_n)$, ($n$ -- positive
integer) may take a specific vector value $y^n=(y_1,\ldots,y_n)$
in $\calY^n$, the $n$--th order Cartesian power of $\calY$, which is
the alphabet of each component of this vector. For $i\le j$
($i$, $j$ -- positive integers), $x_i^j$ will denote the segment
$(x_i,\ldots,x_j)$, where for $i=1$ the subscript will be omitted.


Let $\bu=(u_1,u_2,\ldots)$ be an individual source sequence of symbols in a finite alphabet
$\calU$ of cardinality $|\calU|=J$. 
The sequence $\bu$ is encoded using a general encoder, whose
output at time $t$ is $x_t\in\calX$, where $\calX$ is another finite alphabet\footnote{In 
the general formulations of the joint source--channel coding problem,
the source and the channel are allowed to operate at different rates, and
then, in the case of block codes,
source blocks of a given length may be mapped into channel blocks of a different length. This
degree of freedom, however, is essentially available here too, by redefining $\calU$ and
$\calX$ to be superalphabets of the appropriate sizes.}
of size $|\calX|=K$.
The sequence
$\bx=(x_1,x_2,\ldots)$ is fed into a discrete memoryless channel (DMC), 
characterized by the matrix of single-letter
transition probabilities $\{P(y|x),~x\in\calX,~y\in\calY\}$, where the output
alphabet $\calY$ is
a finite alphabet of size $|\calY|=L$. The channel output
$\by=(y_1,y_2,\ldots)$ is in turn fed into a finite--state decoder, which
is defined by the following recursive equations:
\begin{eqnarray}
v_{t-d}&=&f(z_t,y_t), ~~~~~~~~t=d+1,d+2,\ldots\\
z_{t+1}&=&g(z_t,y_t), ~~~~~~~~t=1,2,\ldots
\end{eqnarray}
where $z_t\in\calZ$ is the decoder state at time $t$, $\calZ$ being a finite
set of states of size $s$,
$v_{t-d}\in\calV$ is the reconstructed sequence, delayed by $d$
time units ($d$ -- positive integer) and $f:\calZ\times\calY\to\calV$
and $g:\calZ\times\calY\to\calZ$ are the output function and
the next--state function, respectively. The reconstruction alphabet $\calV$
of size $M$. 

A slightly more sophisticated model allows the decoder to be equipped
with a modulo--$\ell$ counter, in addition to its state variable.
This means that the functions $f$ and $g$ are allowed to be time--varying
within each period of length $\ell$. In particular, in this case, the decoding equations
would admit the form:
\begin{eqnarray}
\tau&=&t~\mbox{mod}~\ell,~~~~~~~t=1,2,\ldots\\
v_{t-d}&=&f_\tau(z_t,y_t), ~~~~~~~~t=d+1,d+2,\ldots\\
z_{t+1}&=&g_\tau(z_t,y_t), ~~~~~~~~t=1,2,\ldots
\end{eqnarray}

In some applications, one may be interested in a common reconstruction
at both the encoder and decoder (with high probability). In our context, this
means that for a certain positive integer, which we will choose to be $\ell$,
there is a deterministic function $q:\calU^\ell\to\calV^\ell$ such that
\begin{equation}
\label{commonreconstruction}
\lim_{n\to\infty}\frac{\ell}{n}\sum_{i=0}^{n/\ell-1}\mbox{Pr}\{V_{i\ell+1}^{i\ell+\ell}\ne
q(u_{i\ell+1}^{i\ell+\ell})\}=0,
\end{equation}
where here and throughout the sequel, probabilities and expectations are defined
with respect to (w.r.t.) the randomness of the channel. This means that
there is a {target
reconstruction} $\hat{v}^n$, obtained by $n/\ell$ successive applications
of $q(\cdot)$, i.e.,
$\hat{v}_{i\ell+1}^{i\ell+\ell}=q(u_{i\ell+1}^{i\ell+\ell})$,
$i=0,1,2,\ldots,n/\ell-1$, such that $V^n$ is very close to $\hat{v}^n$ in the
sense of eq.\ (\ref{commonreconstruction}). For example, in the traditional
coding theorem of joint source--channel coding, this is achieved by separate
source-- and channel coding, where $\hat{v}_{i\ell+1}^{i\ell+\ell}$ are
rate--distortion reproduction codewords of $u_{i\ell+1}^{i\ell+\ell}$, respectively.

For a given distortion measure $\rho:\calU\times\calV\to \reals$, 
we are interested in deriving lower bounds on the minimum achievable expected
distortion,
$\frac{1}{n}\sum_{t=1}^n\bE\{\rho(u_t,V_t)\}$, as functions of the alphabet
sizes, the number of stated $s$, the allowed delay $d$, and the period $\ell$,
if applicable, with/without a modulo--$\ell$ counter at the decoder, and
with/without the requirement of common reconstruction with high probability.

Throughout our assertions and derivations, we will make heavy use of the following additional
notation. Assume, without essential loss of generality, that 
$\ell$ divide $n$ and consider the segmentation of 
each $n$--vector to $n/\ell$
non--overlapping blocks of length $\ell$, that is,
$$u^n=(\bu_0,\bu_1,\ldots,\bu_{n/\ell-1}),
~~~~~~~~\bu_i=(u_{i\ell+1},u_{i\ell+2},\ldots,u_{i\ell+\ell}),~~~~i=0,1,\ldots,n/\ell-1,$$
and similar definitions for $x^n$, $y^n$, and $v^n$, where
$v_{n-d+1}, v_{n-d+2},\ldots,v_n$ (which are not yet
reconstructed at time $t=n$) are defined as
arbitrary symbols in $\calV$.
Let us define the empirical joint probability mass function
\begin{equation}
\hat{P}_{U^\ell X^\ell Y^\ell V^\ell Z}(u^\ell,x^\ell,y^\ell,v^\ell,z)
=\frac{\ell}{n}\sum_{i=0}^{n/\ell-1}
\calI(\bu_i=u^\ell,\bx_i=x^\ell,\by_i=y^\ell,\bv_i=v^\ell,z_{i\ell+1}=z),
\end{equation}
where $\calI(\cdot)$ is the indicator function of an event.
Correspondingly, unless specified otherwise, $U^\ell$, $X^\ell$, $Y^\ell$,
$V^\ell$ and $Z$ are understood to be random variables jointly distributed according to
$\hat{P}_{U^\ell X^\ell Y^\ell V^\ell Z}$ and all information measures
associated with them will be denoted as in the customary notation conventions of the
information theory literature, but with
``hats'', for example, $\hat{H}(U^\ell)$ is the empirical entropy associated
with $U^\ell$, $\hat{I}(X^\ell;Y^\ell)$ is the empirical mutual information
between $X^\ell$ and $Y^\ell$, and so on. Accordingly, the $\ell$--th order empirical rate
distortion function, associated with $u^n$ and distortion measure $\rho$, is
defined as
\begin{equation}
\hat{R}_{U^\ell}(D)=\min\left\{\frac{1}{\ell}\hat{I}(U^\ell;\tilde{V}^\ell):~
\bE\rho(U^\ell;\tilde{V}^\ell)\le
D\right\},
\end{equation}
where $\tilde{V}^\ell$ is a generic random variable (not to be confused with
$V^\ell$, which is defined empirically), taking on values in
$\calV^\ell$, the mutual information $\hat{I}(U^\ell;\tilde{V}^\ell)$ and
expected distortion $\bE\rho(U^\ell,\tilde{V}^\ell)$ are defined w.r.t.\ 
$\hat{P}_{U^\ell}P_{\tilde{V}^\ell|U^\ell}$, and the minimization is across
all conditional distributions $P_{\tilde{V}^\ell|U^\ell}$.
Here, $\rho(U^\ell,\tilde{V}^\ell)$ is defined additively over the corresponding
components of both vectors.  Similarly, $\hat{D}_{U^\ell}(R)$ is the
corresponding distortion--rate function, which is the inverse of
$\hat{R}_{U^\ell}(D)$, and which is defined as
\begin{equation}
\hat{D}_{U^\ell}(R)=\min\left\{\frac{1}{\ell}\bE\rho(U^\ell,\tilde{V}^\ell):~
\hat{I}(U^\ell;\tilde{V}^\ell)\le\ell
R\right\}.
\end{equation}
In the sequel, we will define some additional empirical rate--distortion
functions and distortion--rate functions, with certain modifications of the
above definitions.

\section{Distortion Bounds Without Common Reconstruction}

We begin from the simpler case where there is no requirement of common
reconstruction. Our first result is the following:

\begin{theorem}
Consider the communication setting described in Section 2.
Let $u^n$ be an individual sequence, let $C$ be the capacity of the
discrete memoryless channel, and let the overall coding--decoding delay be $d$.
Then, for every decoder with $s$ states and a modulo--$\ell$ counter,
\begin{equation}
\frac{1}{n}\sum_{t=1}^n\bE\{\rho(u_t,V_t)\}\ge
\hat{D}_{U^\ell}\left(C+\frac{2\log s+d\log M}{\ell}+\delta_1(\ell,n)\right),
\end{equation}
where 
\begin{equation}
\delta_1(\ell,n)=\frac{(JK)^\ell\log L}{\sqrt{n}}+\frac{(JKL)^\ell\log
e}{2n}+o\left(\frac{1}{\sqrt{n}}\right).
\end{equation}
\end{theorem}
The interesting term, in the argument of the function
$\hat{D}_{U^\ell}(\cdot)$, is the second one, namely, the term $(2\log s+d\log M)/\ell$, which
seemingly plays a role of an effective ``extra capacity'' contributed by the
state variable, that carries memory of past data from block to block and by the allowed delay.
This happens because the lower bound holds for every individual sequence $u^n$
and every encoder and decoder in the allowed class, including ones that happen to
be `tailored' to $u^n$ in a certain sense (for example, the finite--state machine at the decoder
may be designed to periodically produce a certain pattern that happens to be repetitive in
$u^n$). The dependence on $\ell$ is much more complicated, because $\ell$
appears also in the additional term $\delta_1(\ell,n)$, and more importantly,
in the function $\hat{D}_{U^\ell}(\cdot)$ itself. The lower bound is
not necessarily a monotonically decreasing function of $\ell$, but this should
not be surprising since the real optimum performance need not have such a
monotonicity property either. For example, if $u^n$ happens to be periodic (or almost periodic)
with period $\ell$, it seems plausible that it will be reproduced better by a
decoder with a modulo--$\ell$ counter than by one with a modulo--$(\ell+1)$
counter, which obviously cannot keep the synchronization with $u^n$. In the absence of
a modulo--$\ell$ counter at the decoder, Theorem 1 still applies, but then
$\ell$ becomes just a parameter of the bound, with no apparent operative
significance, and since the real distortion,
$\frac{1}{n}\sum_{t=1}^n\bE\{\rho(u_t,V_t)\}$, is then independent of $\ell$,
one may maximize the lower bound w.r.t.\ $\ell$ over a certain set of divisors
of $n$, for which $n/\ell$ is still appreciably large, such that the $o(1/\sqrt{n})$
term would remain negligible.

\vspace{0.3cm}

\noindent
{\it Proof of Theorem 1.}
First, observe that since $\hat{P}_{U^\ell X^\ell Y^\ell V^\ell Z}$ is a legitimate probability
distribution, all the rules of manipulating information measures (the chain
rule, condition reduces entropy, etc.) hold as usual. We will make use of the
fact that 
$\bv_i^{\ell-d}=(v_{i\ell+1},\ldots,v_{i\ell+\ell-d})$ is a deterministic
function of $\by_i$ and $z_{i\ell+1}$
and therefore $(U^\ell,X^\ell)\to Y^\ell\to V^{\ell-d}$ is a Markov chain under
$\hat{P}_{U^\ell X^\ell Y^\ell V^\ell|Z}$, where $V^{\ell-d}$ is random vector formed by
the first $\ell-d$ components of $V^\ell$ (and similarly, below, $V_{\ell-d+1}^\ell$ will denote the
vector formed by the remaining $d$ components).
We then have the following chain of
inequalities 
\begin{eqnarray}
\label{ub}
\hat{I}(U^\ell;V^{\ell-d}|Z)&\le& \hat{I}(U^\ell;Y^\ell|Z)\\
&\le& \hat{I}(U^\ell,X^\ell;Y^\ell|Z)\\
&=&\hat{H}(Y^\ell|Z)-\hat{H}(Y^\ell|U^\ell,X^\ell,Z)\\
&\le&\hat{H}(Y^\ell)-\hat{H}(Y^\ell|U^\ell,X^\ell)+\hat{I}(Z;Y^\ell|U^\ell,X^\ell)\\
&\le&\hat{H}(Y^\ell)-\hat{H}(Y^\ell|U^\ell,X^\ell)+\log s.
\end{eqnarray}
On the other hand,
\begin{eqnarray}
\label{lb}
\hat{I}(U^\ell;V^{\ell-d}|Z)&=& \hat{H}(U^\ell|Z)-\hat{H}(U^\ell|V^{\ell-d},Z)\\
&\ge &\hat{H}(U^\ell)-\hat{I}(Z;U^\ell)-\hat{H}(U^\ell|V^{\ell-d})\\
&\ge &\hat{H}(U^\ell)-\log
s-\hat{H}(U^\ell|V^\ell)-\hat{I}(V_{\ell-d+1}^\ell;U^\ell|V^{\ell-d})\\
&\ge &\hat{I}(U^\ell;V^\ell)-\log s-d\log M,
\end{eqnarray}
and so
\begin{equation}
\hat{I}(U^\ell;V^\ell)\le \hat{H}(Y^\ell)-\hat{H}(Y^\ell|U^\ell,X^\ell)+2\log
s+d\log M.
\end{equation}
Taking now the expectation of both sides, we get
\begin{eqnarray}
\label{dpt}
\bE\hat{I}(U^\ell;V^\ell)&\le&
\bE\hat{H}(Y^\ell)-\bE\hat{H}(Y^\ell|U^\ell,X^\ell)+2\log s+d\log M\nonumber\\
&\le& H(Y^\ell)-\bE\hat{H}(Y^\ell|U^\ell,X^\ell)+2\log s+d\log M
\end{eqnarray}
where in the second line, $H(Y^\ell)$ is the entropy of $Y^\ell$ that is
induced by $\hat{P}_{X^\ell}$ and
the real channel $P_{Y^\ell|X^\ell}$. Here
we have used the fact that 
$\hat{H}(Y^\ell)$ is a concave functional of $\hat{P}_{Y^\ell|X^\ell}$.
As for the evaluation of
$\bE\hat{H}(Y^\ell|U^\ell,X^\ell)$, we invoke
the following result (see \cite{Atteson99}, \cite{CB90} and [19, Proposition
5.2] therein, as well as \cite[Appendix A]{MW04}): 
Let $\hat{P}_n$ be the first order
empirical distribution associated with an $n$--sequence drawn from a
memoryless $m$--ary source $P$. Then,
\begin{equation}
n\cdot\bE D(\hat{P}_n\|P)=\frac{(m-1)\log e}{2}+o(1),
\end{equation}
which is equivalent to
\begin{equation}
\bE \hat{H} = H-\frac{(m-1)\log e}{2n}-o\left(\frac{1}{n}\right),
\end{equation}
where $\hat{H}$ is the corresponding empirical entropy and $H$ is the true
entropy. We now apply this result to the `source'
$P(y^\ell|u^\ell,x^\ell)\equiv P(y^\ell|x^\ell)$ for every pair
$(u^\ell,x^\ell)$ that appears more than $\epsilon n/\ell$ times as
$\ell$--blocks along the (deterministic) sequence pair
$(u^n,x^n)$.
\begin{eqnarray}
&&\bE\hat{H}(Y^\ell|U^\ell,X^\ell)\\
&=&\bE\left\{\sum_{u^\ell,x^\ell}\hat{P}_{U^\ell
X^\ell}(u^\ell,x^\ell)\hat{H}(Y^\ell|U^\ell=u^\ell,X^\ell=x^\ell)\right\}\\
&\ge&\sum_{\{u^\ell,x^\ell:~\hat{P}_{U^\ell X^\ell}(u^\ell,x^\ell)\ge\epsilon\}}\hat{P}_{U^\ell
X^\ell}(u^\ell,x^\ell)\bE\hat{H}(Y^\ell|U^\ell=u^\ell,X^\ell=x^\ell)\\
&=&\sum_{\{u^\ell,x^\ell:~\hat{P}_{U^\ell X^\ell}(u^\ell,x^\ell)\ge\epsilon\}}\hat{P}_{U^\ell
X^\ell}(u^\ell,x^\ell)\left[H(Y^\ell|X^\ell=x^\ell)-\frac{(L^\ell-1)\log
e}{2n\hat{P}_{U^\ell
X^\ell}(u^\ell,x^\ell)/\ell}-o\left(\frac{\ell}{n\epsilon}\right)\right]\\
&\ge&\sum_{\{u^\ell,x^\ell:~\hat{P}_{U^\ell X^\ell}(u^\ell,x^\ell)\ge\epsilon\}}\hat{P}_{U^\ell
X^\ell}(u^\ell,x^\ell)H(Y^\ell|X^\ell=x^\ell)-\frac{\ell(JKL)^\ell
\log e}{2n}
-o\left(\frac{\ell}{n\epsilon}\right)\\
&\ge&\sum_{u^\ell,x^\ell}\hat{P}_{U^\ell
X^\ell}(u^\ell,x^\ell)H(Y^\ell|X^\ell=x^\ell)-
\sum_{\{u^\ell,x^\ell:~\hat{P}_{U^\ell,X^\ell}(u^\ell,x^\ell)<\epsilon\}}\hat{P}_{U^\ell
X^\ell}(u^\ell,x^\ell)H(Y^\ell|X^\ell=x^\ell)\nonumber\\
& &-\frac{\ell (JKL)^\ell
\log e}{2n}-
o\left(\frac{\ell}{n\epsilon}\right)\\
&\ge&
H(Y^\ell|X^\ell)-\epsilon (JK)^\ell\cdot \ell\log L
-\frac{\ell (JKL)^\ell
\log e}{2n}-
o\left(\frac{\ell}{n\epsilon}\right)\\
&=&H(Y^\ell|X^\ell)-\ell\cdot \delta_0(\epsilon,\ell,n),
\end{eqnarray}
where we have defined
\begin{equation}
\delta_0(\epsilon,\ell,n)=\epsilon (JK)^\ell\log L+\frac{(JKL)^\ell\log
e}{2n}+o\left(\frac{1}{n\epsilon}\right).
\end{equation}
Taking $\epsilon=1/\sqrt{n}$, we define:
\begin{equation}
\delta_1(\ell,n)=\delta_0\left(\frac{1}{\sqrt{n}},\ell,n\right)=
\frac{(JK)^\ell\log L}{\sqrt{n}}+\frac{(JKL)^\ell\log
e}{2n}+o\left(\frac{1}{\sqrt{n}}\right).
\end{equation}
On substituting the inequality
\begin{equation}
\bE\hat{H}(Y^\ell|U^\ell,X^\ell)\ge H(Y^\ell|X^\ell)-\ell\delta_1(\ell,n)
\end{equation}
into eq.\ (\ref{dpt}), we get
\begin{eqnarray}
\bE\hat{I}(U^\ell;V^\ell)&\le& I(X^\ell;Y^\ell)+
2\log s+d\log M+\ell\delta_1(\ell,n)\\
&\le& \ell C+
2\log s+d\log M+\ell\delta_1(\ell,n).
\end{eqnarray}
Now, denoting by $\hat{\bE}$ the empirical expectation
(w.r.t.\ $\hat{P}_{U^\ell X^\ell Y^\ell V^\ell Z}$), we
obviously have
\begin{eqnarray}
\bE\hat{I}(U^\ell;V^\ell)&\ge&\ell\cdot
\bE\hat{R}_{U^\ell}\left(\frac{1}{\ell}\hat{\bE}\rho(U^\ell,V^\ell)\right)\\
&=&\ell\cdot
\bE\hat{R}_{U^\ell}\left(\frac{1}{n}\sum_{t=1}^n\rho(u_t,V_t)\right)\\
&\ge&\ell\cdot
\hat{R}_{U^\ell}\left(\frac{1}{n}\sum_{t=1}^n\bE\rho(u_t,V_t)\right),
\end{eqnarray}
where in the last line, we have used the convexity of the rate--distortion
function. Finally, we get
\begin{equation}
\hat{R}_{U^\ell}\left(\frac{1}{n}\sum_{t=1}^n\bE\rho(u_t,V_t)\right)\le
C+\frac{2\log s+d\log M}{\ell}+\delta_1(\ell,n),
\end{equation}
or 
\begin{equation}
\frac{1}{n}\sum_{t=1}^n\bE\rho(u_t,V_t)\ge
\hat{D}_{U^\ell}\left(C+\frac{2\log s+d\log M}{\ell}+\delta_1(\ell,n)\right).
\end{equation}
This completes the proof of Theorem 1. $\Box$\\

While the lower bound of Theorem 1 is not quite explicit (primarily
because of the complicated dependence of the function
$\hat{D}_{U^\ell}(\cdot)$ on $\ell$ when $u^n$ is arbitrary), 
we next propose an alternative lower
bound, which is simpler and more explicit. The price of this simplicity, however, is a
possible loss of tightness, The idea is based on the Shannon lower bound.
Suppose that $\calU=\calV$ is a group 
and the distortion measure $\rho(u,v)$ depends only on the difference
$u-v$ for a well defined subtraction operation on the group 
(e.g., subtraction modulo $J$). Accordingly, we denote
$\rho(u,v)=\varrho(v-u)$. We define the function $\Phi(D)$ to be the maximum
entropy of a random variable $W$ over an alphabet of size $J$, subject to the
constraint $\bE\varrho(W)\le D$. We also define
\begin{equation}
\Psi(x)=\left\{\begin{array}{ll}
0 & x<0\\
\Phi^{-1}(x) & x\ge 0\end{array}\right.
\end{equation}
Then, our next result is the following.

\begin{theorem}
Consider the communication setting described in Section 2.
Let $u^n$ be an individual sequence, let $C$ be the capacity of the
discrete memoryless channel, and let the overall coding--decoding delay be
$d$. Then, for every decoder with $s$ states and a modulo--$\ell$ counter,
\begin{equation}
\frac{1}{n}\sum_{t=1}^n\bE\{\varrho(V_t-u_t)\}\ge
\Psi\left(\frac{c(u^n)\log c(u^n)}{n}-C-
\frac{2\log s+d\log M}{\ell}-\delta_2(\ell,n)\right),
\end{equation}
where $c(u^n)$ is the number of phrases associated with incremental parsing
\cite{ZL78} of $u^n$ and
\begin{equation}
\delta_2(\ell,n)=\delta_1(\ell,n)+\frac{2\ell(1+\log J)^2}{(1-\epsilon_n)\log
n}+\frac{2\ell J^{2\ell}\log J}{n}+\frac{1}{\ell},
\end{equation}
$\epsilon_n$ being a positive sequence tending to zero as $n\to\infty$.
\end{theorem}

An important feature
of this bound is that the dependence on $\ell$ is now fairly explicit as it 
appears only in the expression $\delta_2(\ell,n)+(2\log s+d\log M)/\ell$, and so, the effect of the choice of $\ell$
can be better understood. Indeed, for decoders that are not equipped
with a counter, the maximization of the bound over $\ell$, which is
equivalent to the minimization of $\delta_2(\ell,n)+(2\log s+d\log M)/\ell$, is easier now. In
particular, it is
clear that $\ell$ should be $o(\log n)$ for this expression to vanish as
$n\to\infty$. Another
interesting point here is that the bound depends on $u^n$ only via its
Lempel-Ziv complexity, $c(u^n)\log c(u^n)/n$. This is not a trivial fact, because
the Lempel--Ziv complexity refers to the compressibility of $u^n$ using finite--state
encoders, whereas here, the encoder is not limited to be a finite--state
machine -- only the decoder has such a limitation. 

\vspace{0.3cm}

\noindent
{\it Proof of Theorem 2.}
Defining
$V^\ell-U^\ell$ as the component-wise difference between the two vectors,
we have:
\begin{eqnarray}
\ell\cdot
\hat{R}_{U^\ell}(D)&=&\hat{H}(U^\ell)-\max\{H(U^\ell|V^\ell):~\bE\varrho(V^\ell-U^\ell
)\le \ell D\}\\
&=&\hat{H}(U^\ell)-\max\{H(V^\ell-U^\ell|V^\ell):~\bE\varrho(V^\ell-U^\ell
)\le \ell D\}\\
&=&\hat{H}(U^\ell)-\max\{H(W^\ell|V^\ell):~\bE\varrho(W^\ell)\le\ell D\}\\
&\ge&\hat{H}(U^\ell)-\max\{H(W^\ell):~\bE\varrho(W^\ell)\le\ell D\}\\
&\ge&\hat{H}(U^\ell)-
\max\left\{\sum_{i=1}^\ell H(W_i):~\sum_{i=1}^\ell\bE\varrho(W_i)\le\ell
D\right\}\\
&\ge&\hat{H}(U^\ell)-\max\left\{\sum_{i=1}^\ell\Phi(\bE\varrho(W_i)):~
\sum_{i=1}^\ell\bE\varrho(W_i)\le\ell D\right\}\\
&\ge&\hat{H}(U^\ell)-
\max\left\{\ell\cdot\Phi\left(\frac{1}{\ell}\sum_{i=1}^\ell\bE\varrho(W_i)\right):~
\sum_{i=1}^\ell\bE\varrho(W_i)\le\ell D\right\}\\
&=&\hat{H}(U^\ell)-\ell\cdot \Phi(D),
\end{eqnarray}
where in the last two lines, we have used concavity and the monotonicity of
$\Phi(\cdot)$, respectively.
Now, it is shown in \cite[eq.\ (21)]{Merhav13} (see also \cite{Merhav00}) that
\begin{equation}
\hat{H}(U^\ell)\ge \ell\cdot\left[\frac{c(u^n)\log
c(u^n)}{n}-\delta(\ell,n)\right],
\end{equation}
where 
\begin{equation}
\delta(\ell,n)=\frac{2\ell(1+\log J)^2}{(1-\epsilon_n)\log n}+\frac{2\ell
J^{2\ell}\log J}{n}+\frac{1}{\ell},
\end{equation}
$\epsilon_n$ being a positive sequence tending to zero,
and $c(u^n)$ is the number of phrases in $u^n$ resulting from Lempel--Ziv
incremental parsing.
Thus,
\begin{eqnarray}
\bE\hat{R}_{U^\ell}\left(\frac{1}{n}\sum-{t=1}^n\rho(V_t-u_t)\right)&\ge&\frac{c(u^n)\log
c(u^n)}{n}-\bE\Phi\left(\frac{1}{n}\sum_{t=1}^n
\varrho(V_t-u_t)\right)-\delta(\ell,n)\\
&\ge&\frac{c(u^n)\log
c(u^n)}{n}-\Phi\left(\frac{1}{n}\sum_{t=1}^n\bE\varrho(V_t-u_t)\right)-
\delta(\ell,n).
\end{eqnarray}
and we end up with
\begin{equation}
\Phi\left(\frac{1}{n}\sum_{t=1}^n\bE\varrho(V_t-u_t)\right)\ge
\frac{c(u^n)\log c(u^n)}{n}-C-\frac{2\log s+d\log
M}{\ell}-\delta_2(\ell,n)
\end{equation}
or
\begin{equation}
\frac{1}{n}\sum_{t=1}^n\bE\varrho(V_t-u_t)\ge
\Psi\left(\frac{c(u^n)\log c(u^n)}{n}-C-\frac{2\log s+d\log M}{\ell}-\delta_2(\ell,n)\right).
\end{equation}
This completes the proof of Theorem 2. $\Box$

\section{Distortion Bounds Under Common Reconstruction}

Consider next the case where, in addition to the above--mentioned 
limitations on the decoder,
an additional constraint is imposed, which is the constraint of almost
deterministic reconstruction at the level of $\ell$--blocks.
This setting is formalized
as follows. For a given vanishing sequence $\epsilon_n\in[0,1]$, we insist that
\begin{equation}
\label{cr}
\bE\hat{\mbox{Pr}}\{V^\ell\ne \hat{V}^\ell\}
\equiv
\frac{\ell}{n}\sum_{i=0}^{n/\ell-1}\mbox{Pr}\{V_{i\ell+1}^{i\ell+\ell}\ne
\hat{v}_{i\ell+1}^{i\ell+\ell}\}
\le\epsilon_n, 
\end{equation}
where
$\hat{V}^\ell=q(U^\ell)$ (and
$\hat{v}_{i\ell+1}^{i\ell+\ell}=q(u_{i\ell+1}^{i\ell+\ell})$), for some deterministic function $q$, is the target
reconstruction. We will assume, in this section, that
$\rho_{\max}\dfn\max_{u,v}\rho(u,v) < \infty$.
Our lower bound for this case is given by the following
theorem.

\begin{theorem}
Consider the communication setting described in Section 2.
Let $u^n$ be an individual sequence, let $C$ be the capacity of the
discrete memoryless channel, and let the overall coding--decoding delay be
$d$. Then, for every decoder with $s$ states, a modulo--$\ell$ counter and
a common reconstruction constraint defined as in eq.\ (\ref{cr}):
\begin{equation}
\frac{1}{n}\sum_{t=1}^n\bE\{\rho(u_t,V_t)\}\ge
\tilde{D}_{U^\ell}\left(C+\frac{2\log s+d\log
M}{\ell}+\delta_2(\ell,n)+2\Delta(\epsilon_n)\right)-\rho_{\max}\epsilon_n,
\end{equation}
where $\Delta(\epsilon_n)=h_2(\epsilon_n)+\epsilon_n\ell\log J$, $h_2(\cdot)$
being the binary entropy function, and
\begin{equation}
\tilde{D}_{U^\ell}(R)=\min_q\left\{\frac{1}{\ell}\hat{\bE}\rho(U^\ell,q(U^\ell)):~\hat{H}(q(U^\ell))\le
\ell R\right\}.
\end{equation}
\end{theorem}

\noindent
{\it Proof of Theorem 3.}
First, under the assumption of common reconstruction (\ref{cr}), one readily
finds, using Fano's inequality, that
\begin{equation}
\bE\hat{H}(V^\ell|U^\ell)\le \Delta(\epsilon_n),
\end{equation}
where the concavity of the function $\Delta(\cdot)$ was used in order to
insert the expectation into the argument of this function in order to get the
real probability of error.
Thus,
\begin{eqnarray}
\bE\hat{I}(U^\ell;V^\ell)&=&\bE\hat{H}(V^\ell)-\bE\hat{H}(V^\ell|U^\ell)\\
&\ge&\bE\hat{H}(V^\ell)-\Delta(\epsilon_n).
\end{eqnarray}
Now,
\begin{eqnarray}
\bE\hat{H}(V^\ell)&\ge&\hat{H}(\hat{V}^\ell)-\bE\hat{H}(\hat{V}^\ell|V^\ell)\\
&\ge&\hat{H}(\hat{V}^\ell)-\Delta(\epsilon_n)
\end{eqnarray}
and so,
\begin{equation}
\bE\hat{I}(U^\ell;V^\ell)\ge \hat{H}(\hat{V}^\ell)-2\Delta(\epsilon_n).
\end{equation}
Now, observe that
\begin{eqnarray}
\frac{1}{n}\sum_{t=1}^n\rho(u_t,\hat{v}_t)&=&
\hat{\bE}\left\{\frac{1}{\ell}\rho(U^\ell,q(U^\ell))\right\}\\
&=&\frac{1}{\ell}\sum_{\{(u^\ell,v^\ell):~q(u^\ell)=v^\ell\}}
\hat{P}_{U^\ell,V^\ell}(u^\ell,v^\ell)\rho(u^\ell,v^\ell)+\nonumber\\
&&\frac{1}{\ell}\sum_{\{(u^\ell,v^\ell):~q(u^\ell)\ne v^\ell\}}
\hat{P}_{U^\ell,V^\ell}(u^\ell,v^\ell)\rho(u^\ell,q(u^\ell))\\
&\le&\frac{1}{\ell}\sum_{u^\ell,v^\ell}
\hat{P}_{U^\ell,V^\ell}(u^\ell,v^\ell)\rho(u^\ell,v^\ell)+
\rho_{\max}\cdot\sum_{\{(u^\ell,v^\ell):~q(u^\ell)\ne v^\ell\}}
\hat{P}_{U^\ell,V^\ell}(u^\ell,v^\ell)\\
&=&\frac{1}{n}\sum_{t=1}^n
\rho(u_t,V_t)+\rho_{\max}\cdot\hat{\mbox{Pr}}\{V^\ell\ne\hat{V}^\ell\}
\end{eqnarray}
and so, taking the expectation of both sides, we get
\begin{equation}
\frac{1}{n}\sum_{t=1}^n\rho(u_t,\hat{v}_t)\le \frac{1}{n}\sum_{t=1}^n
\bE\rho(u_t,V_t)+\rho_{\max}\epsilon_n.
\end{equation}
Thus, defining
\begin{equation}
\tilde{R}_{U^\ell}(D)=\min_q\{\frac{1}{\ell}\hat{H}(q(U^\ell)):~\hat{\bE}\rho(U^\ell,q(U^\ell))\le\ell
D\},
\end{equation}
we readily have
\begin{eqnarray}
\bE\hat{I}(U^\ell;V^\ell)&\ge& \hat{H}(q(U^\ell))-2\Delta(\epsilon_n)\\
&\ge& \ell\tilde{R}_{U^\ell}\left(\frac{1}{n}\sum_{t=1}^n\rho(u_t,\hat{v}_t)\right)-2\Delta(\epsilon_n)\\
&\ge&\ell\tilde{R}_{U^\ell}\left(\frac{1}{n}\sum_{t=1}^n\bE\rho(u_t,V_t)+\rho_{\max}\epsilon_n\right)-
2\Delta(\epsilon_n).
\end{eqnarray}
This means, of course, that
\begin{equation}
\frac{1}{n}\sum_{t=1}^n\bE\rho(u_t,V_t)\ge
\tilde{D}_{U^\ell}\left(C+\frac{2\log s+d\log
M}{\ell}+\delta_2(\ell,n)+\frac{2\Delta(\epsilon_n)}{\ell}\right)-\rho_{\max}\epsilon_n,
\end{equation}
completing the proof of Theorem 3. $\Box$

Here too, performance can be expressed in terms of Lempel--Ziv complexity,
as $\hat{H}(q(U^\ell))/\ell\ge [c(\hat{v}^n)\log
c(\hat{v}^n)]/n-\delta'(\ell,n)$, where $\delta'(\ell,n)$ is defined just
like $\delta(\ell,n)$, but with $J$ replaced by $M$. Thus,
\begin{equation}
\bE\hat{I}(U^\ell;V^\ell)\ge\ell
R_{LZ}\left(\frac{1}{n}\sum_{t=1}^n\bE\rho(u_t,V_t)+
\rho_{\max}\epsilon_n\bigg| u^n\right)-2\Delta(\epsilon_n)-\delta'(\ell,n),
\end{equation}
where
\begin{eqnarray}
R_{LZ}(D|u^n)&=&\min_q\left\{\frac{c(\hat{v}^n)\log c(\hat{v}^n)}{n}:
\hat{v}_{i\ell+1}^{i\ell+\ell}=q(u_{i\ell+1}^{i\ell+\ell}),\right.\nonumber\\
& &\left. i=0,1,\ldots,n/\ell-1,~
\frac{1}{n}\sum_{i=1}^n\rho(u_t,\hat{v}_t)\le D\right\}.
\end{eqnarray}
Note that in Section 3, we were able to get bounds on the expected distortion,
thanks to the convexity of
$\hat{R}_{U^\ell}(\cdot)$ and the concavity of $\Phi(\cdot)$, whereas now,
we obtained such a bound by using the proximity between
the actual expected distortion and the distortion between $u^n$ and its
intended reconstruction $\hat{v}^n$.

\section{Linear Encoders and Decoders}

So far, we have dealt with finite alphabets only. It is possible to derive
analogous results for continuous alphabets, if the encoder and decoder are
limited to be linear. In this section, we provide a brief outline how this can
be done, by presenting a parallel result to the Theorem 1.

Consider the following structure: The encoder 
is given by
\begin{equation}
x_t=\sum_{i=1}^\infty a_ix_{t-i}+
\sum_{i=0}^\infty b_iu_{t-i},
\end{equation}
where $\{a_i\}$ and $\{b_i\}$ are real--valued parameters, chosen such that the encoder would satisfy a
certain input constraint.
The finite--state decoder we had before\footnote{For simplicity, we now refer
to the one without the modulo-$\ell$ counter.} is replaced by a decoder with the
same structure, except that now $f$ and $g$ are linear functions (i.e., state--space
representation):
\begin{eqnarray}
v_{t-d}&=&\alpha z_t+\beta y_t\\
z_{t+1}&=&\gamma z_t+\delta y_t.
\end{eqnarray}
We will assume, for the sake of simplicity, that $u_t$, $x_t$, $y_t$, $v_t$
and $z_t$ are all real--valued variables (scalars), although our discussion can be
generalized to the vector case ($u_t,v_t\in\reals^k$, $x_t,y_t\in\reals^m$,
$z_t\in\reals^p$, $k$, $m$ and $p$ positive integers), 
in which case, $\{a_i\}$, $\{b_i\}$, $\alpha$, $\beta$, 
$\gamma$ and $\delta$ become matrices of the corresponding dimensions.
The channel is assumed to be a
discrete--time AWGN, i.e., $Y_t=x_t+N_t$, where $N_t$ is a stationary, i.i.d.\
zero--mean Gaussian process with variance $\sigma^2$.

Consider first\footnote{This assumption will be dropped soon.} the case where $\{u_t\}$ is a zero--mean, stationary Gaussian
process, independent of $\{N_t\}$, and so, its
notation is temporarily changed to $\{U_t\}$. Consequently, all other signals in
the system become random processes, and accordingly, their notation here will use capital
letters. Due to
the linearity of the
systems, $\{(U_t,X_t,Y_t,V_t,Z_t),~-\infty < t < \infty\}$ are jointly Gaussian processes.
We assume that these processes are jointly stationary.
We also assume that the system is non--degenerated\footnote{For example, if
$\gamma=\delta=0$ and hence $Z_t\equiv 0$, or if $\alpha=\beta=0$ and hence
$V_t\equiv 0$, the system is obviously degenerated.}
in the sense that
\begin{equation}
\epsilon_Z^2\dfn \lim_{n\to\infty} \mbox{mmse}\{Z_1|U_1^n,X_1^n,Y_1^n\} > 0
\end{equation}
and similarly
\begin{equation}
\epsilon_V^2\dfn \lim_{n\to\infty} \mbox{mmse}\{V_0|V_{-n}^{-1},U_{-n}^d\} >
0,
\end{equation}
where $\mbox{mmse}\{A|B\}=\bE[A-\bE(A|B)]^2$ designates the minimum mean
squared error in estimating a random variable $A$ from another random variable
$B$, and where the limits obviously exist due to the non--increasing monotonicity of 
$\mbox{mmse}\{Z_1|U_1^n,X_1^n,Y_1^n\}$ and
$\mbox{mmse}\{V_0|V_{-n}^{-1},U_{-n}^d\}$ as functions on $n$. The parameters
$\epsilon_Z^2$ and $\epsilon_V^2$ are constants that depend on the
auto-correlation function of the source, on the noise variance of noise,
$\sigma^2$, and on the parameters of the encoder and decoder,
$\{a_i\}$, $\{b_i\}$, $\alpha$, $\beta$, 
$\gamma$ and $\delta$.
Obviously, $\epsilon_Z^2\le \sigma_Z^2$ and
$\epsilon_V^2\le \sigma_V^2$, where $\sigma_Z^2$ and $\sigma_V^2$ are the
variances of $Z_t$ and $V_t$, respectively.
We define $U^\ell=(U_1,\ldots,U_\ell)$,
$X^\ell=(X_1,\ldots,X_\ell)$,
$Y^\ell=(Y_1,\ldots,Y_\ell)$,
$V^\ell=(V_1,\ldots,V_\ell)$, and $Z=Z_1$.
We begin similarly as in eqs.\ (\ref{ub}), but the last step must be modified
slightly:
\begin{eqnarray}
\label{ubl}
I(U^\ell;V^{\ell-d}|Z)&\le& I(U^\ell;Y^\ell|Z)\\
&\le& I(U^\ell,X^\ell;Y^\ell|Z)\\
&=&h(Y^\ell|Z)-h(Y^\ell|U^\ell,X^\ell,Z)\\
&\le&h(Y^\ell)-h(Y^\ell|U^\ell,X^\ell)+I(Z;Y^\ell|U^\ell,X^\ell)\\
&=&h(Y^\ell)-h(Y^\ell|X^\ell)+\frac{1}{2}\log\frac{\mbox{mmse}\{Z|U^\ell,X^\ell\}}
{\mbox{mmse}\{Z|U^\ell,X^\ell,Y^\ell\}}\\
&\le&I(X^\ell;Y^\ell)+\frac{1}{2}\log\frac{\sigma_Z^2}
{\epsilon_Z^2}\\
&\le&\ell C+\frac{1}{2}\log\frac{\sigma_Z^2}
{\epsilon_Z^2}.
\end{eqnarray}
On the other hand,
\begin{eqnarray}
I(U^\ell;V^{\ell-d}|Z)&=&
h(U^\ell|Z)-h(U^\ell|V^{\ell-d},Z)\\
&\ge&h(U^\ell|Z)-h(U^\ell|V^{\ell-d})\\
&=&h(U^\ell)-I(Z;U^\ell)-h(U^\ell|V^\ell)-I(V_{\ell-d+1}^\ell;U^\ell|V^{\ell-d})\\
&=&h(U^\ell)-h(U^\ell|V^\ell)-I(Z;U^\ell)-\sum_{i=\ell-d+1}^\ell
I(V_i;U^\ell|V^{i-1})\\
&\ge&I(U^\ell;V^\ell)-\frac{1}{2}\log\frac{\sigma_Z^2}{\epsilon_Z^2}-
\frac{d}{2}\log\frac{\sigma_V^2}{\epsilon_V^2},
\end{eqnarray}
and so,
\begin{equation}
I(U^\ell;V^\ell)\le \ell C+\log
\frac{\sigma_Z^2}{\epsilon_Z^2}+\frac{d}{2}\log\frac{\sigma_V^2}{\epsilon_V^2}.
\end{equation}
This is quite analogous to the bounds we obtained in the finite--alphabet
case, but now $\log s$ and $\log M$ are replaced by $\log
\frac{\sigma_Z}{\epsilon_Z}$ and $\log\frac{\sigma_V}{\epsilon_V}$,
respectively, thus $\frac{\sigma_Z}{\epsilon_Z}$ and
$\frac{\sigma_V}{\epsilon_V}$ play roles of effective alphabet sizes
(or effective resolution levels) of the variables $Z_t$ and $V_t$,
respectively. Now, clearly,
in the Gaussian case, $I(U^\ell;V^\ell)$ depends on the joint density of
$(U^\ell,V^\ell)$ only via the covariance matrix of this random vector.
Equivalently, consider the class of Gaussian channels from $U^\ell$ to
$V^\ell$, defined by
\begin{equation}
V^\ell=GU^\ell+W^\ell
\end{equation}
where $G$ is a deterministic $\ell\times\ell$ matrix and $W^\ell$ is a
zero--mean
Gaussian vector, independent of $U^\ell$, with covariance matrix $\Sigma_W$.
Denoting the covariance matrix of $U^\ell$ by $\Sigma_U$, then
\begin{equation}
I(U^\ell;V^\ell)=\frac{1}{2}\log
\frac{\mbox{det}(G\Sigma_UG^T+\Sigma_W)}{\mbox{det}(\Sigma_W)}=
\frac{1}{2}\log\mbox{det}(I+\Sigma_W^{-1}G\Sigma_UG^T)
\end{equation}
Thus, defining
\begin{equation}
R_\ell(D)=\min_{G,\Sigma_W}\left\{\frac{1}{2\ell}\log\mbox{det}(I+\Sigma_W^{-1}G\Sigma_UG^T):~
\bE\rho(U^\ell,V^\ell)\le\ell D\right\},
\end{equation}
where $\rho$ designates the quadratic distortion measure (or any other
distortion measure that such that $\bE\rho(U^\ell,V^\ell)$ depends only on the
covariance matrix of $(U^\ell,V^\ell)$),
we have
\begin{equation}
\label{dptl}
R_\ell(D)\le  C+
\frac{1}{\ell}\log\frac{\sigma_Z^2}{\epsilon_Z^2}+
\frac{d}{2\ell}\log\frac{\sigma_V^2}{\epsilon_V^2},
\end{equation}
or 
\begin{equation}
\frac{1}{n}\sum_{t=1}^n\bE\rho(U_t,V_t)\ge D_\ell\left(C+
\frac{1}{\ell}\log\frac{\sigma_Z^2}{\epsilon_Z^2}+
\frac{d}{2\ell}\log\frac{\sigma_V^2}{\epsilon_V^2}\right),
\end{equation}
Now, the l.h.s.\ of (\ref{dptl}) depends only on the covariance matrix of
the source, whereas $C$ (or $I(X^\ell;Y^\ell)$) depends only on the covariance
matrix $\Sigma_X$ of $X^\ell$ and the covariance matrix $\Sigma_N$ of the noise
vector, which we have taken to be $\sigma^2I$. Since the encoding and decoding systems are linear,
the auto-correlation cross--correlation functions of their outputs depend only on those of their
inputs (for a given linear encoder and decoder), no matter whether these
processes are Gaussian or not. The expected distortion also depends on the
joint density of $U^\ell,V^\ell)$ only via the variances and covariances of
their components.
Consequently, at this point, the Gaussian assumption becomes immaterial.
The source $U^\ell$ may have any pdf with a given covariance matrix $\Sigma_U$.
In particular, we can take $\Sigma_U$ to be the empirical covariance matrix of
a deterministic source sequence $u^n$. In this case, in the above chains of inequalities,
all information measures should be replaced by their empirical counterparts,
which depend on the empirical covariances instead of the true covariances.
The only exception is that, similarly as in the finite alphabet case, in eq.\
(\ref{ubl}), it is no longer true that $\hat{h}(Y^\ell|X^\ell,U^\ell)=
\hat{h}(Y^\ell|X^\ell)$, since there might be empirical correlations between
the source vector and the noise vector. However,
$\bE\hat{h}(Y^\ell|X^\ell,U^\ell)$ tends to $h(Y^\ell|X^\ell)$ by the weak law
of large numbers, so as before,
upon taking expectations, one can obtain a distortion bound analogous to the one
we obtained in the finite--alphabet case. In particular,
for the quadratic distortion measure, we have:
\begin{eqnarray}
\frac{1}{n}\sum_{t=1}^n\bE\rho(u_t,V_t)&\ge& 
\min_{G,\Sigma_W}\left\{\frac{1}{n}\sum_{i=0}^{n/\ell-1}
\bE\rho(u_{i\ell+1}^{i\ell+\ell},Gu_{i\ell+1}^{i\ell+\ell}+W_{i\ell+1}^{i\ell+\ell}):~\right.\nonumber\\
& &\left.\frac{1}{2\ell}\log \mbox{det}(I+\Sigma_W^{-1}G\hat{\Sigma}_UG^T)\le
C+\frac{1}{\ell}\log\frac{\sigma_Z^2}{\epsilon_Z^2}+
\frac{d}{2\ell}\log\frac{\sigma_V^2}{\epsilon_V^2}+\epsilon_n\right\}\\
&=&\min_{G,\Sigma_W}\left\{
\mbox{tr}\{(G-I)\hat{\Sigma}_U(G^T-I)+\frac{1}{\ell}\Sigma_W\}:~\right.\nonumber\\
& &\left.\frac{1}{2\ell}\log \mbox{det}(I+\Sigma_W^{-1}G\hat{\Sigma}_UG^T)\le
C+\frac{1}{\ell}\log\frac{\sigma_Z^2}{\epsilon_Z^2}+
\frac{d}{2\ell}\log\frac{\sigma_V^2}{\epsilon_V^2}+\epsilon_n\right\},
\end{eqnarray}
where $\hat{\Sigma}_U=\frac{\ell}{n}\sum_{i=0}^{n/\ell-1}\bu_i\bu_i^T$ is the
empirical covariance of the source,
$W_{i\ell+1}^{i\ell+\ell}$ is a zero--mean random vector with
covariance matrix $\Sigma_W$ and $\epsilon_n$ is the vanishing difference
between $\bE\hat{h}(Y^\ell|X^\ell,U^\ell)/\ell$ and $h(Y|X)$.
The point here is that for the purpose of obtaining a lower bound on the distortion
attainable by linear encoders and decoders,
we are replacing the optimization over infinitely many
parameters $\{a_i\}$, $\{b_i\}$, $\alpha$, $\beta$, $\gamma$, and $\delta$, by
optimization over two $\ell\times\ell$ matrices, $G$ and $\Sigma_W$, at the
possible rate loss of $\frac{1}{\ell}\log\frac{\sigma_Z^2}{\epsilon_Z^2}+
\frac{d}{2\ell}\log\frac{\sigma_V^2}{\epsilon_V^2}+\epsilon_n$, which vanishes
as $\ell$ and $n$ grow.
Thus, the parameter $\ell$ trades off the
quality of the bound (its tightness) with the complexity of the optimization.

Note that here our bounds are a bit weaker than in the
finite--alphabet case, in the sense that they depend on the
competing linear system with parameters $\{a_i\}$, $\{b_i\}$, $\alpha$, $\beta$,
$\gamma$ and $\delta$ (via $\epsilon_V^2$ and $\epsilon_Z^2$). However, the
dependence on these parameters becomes weaker and weaker as $\ell$ grows
without bound.


\section*{Appendix}
\renewcommand{\theequation}{A.\arabic{equation}}
    \setcounter{equation}{0}

{\it Some Concerns About the Proof of Theorem 3 in \cite{Ziv80}.}

First, it should be pointed out that
in \cite[p.\ 140]{Ziv80}, the encoder was also assumed to be a finite--state
machine, and so, in this appendix, following the notation of \cite{Ziv80},
the state of the encoder is denoted by $z_t$
and the state of the decoder is denoted by $z_t'$.

In \cite{Ziv80}, the joint probability distribution of all random
variables was defined (in our notation) to be
\begin{eqnarray}
& &\hat{P}_{U^\ell X^\ell Y^\ell V^\ell ZZ'}(u^\ell,x^\ell,y^\ell,v^\ell,
z,z')\nonumber\\
&=&P(z,z')\hat{P}_{U^\ell}(u^\ell)\hat{P}_{X^\ell|U^\ell,Z}(x^\ell|u^\ell,z)
P(y^\ell|x^\ell)\hat{P}_{V^\ell|X^\ell,Z'}(v^\ell|x^\ell,z'),
\end{eqnarray}
where $P(z,z')$ is the expectation of the joint empirical distribution of the
state of the encoder, denoted here by $Z$,
and the state of the decoder, denoted here by $Z'$, at the
beginnings
of all $\ell$-blocks, and $P(y^\ell|x^\ell)$ is the {\it real} conditional
probability
associated with the channel. First, observe that according to this
definition, $U^\ell$ is taken to be independent of $Z$ and $Z'$, which is inconsistent
with the fact that
the encoder state $Z$ varies in response to the source
and that there might be empirical dependencies between successive $\ell$--blocks
of the source. Also, according to
this definition, $Y^\ell$ is independent of $Z'$ given $X^\ell$, which
similarly to the earlier comment, does
not seem to settle with the fact that $Z'$ responds to the decoder input
$Y^\ell$.

Another issue is the use of the data processing theorem when it
comes to empirical distributions. For example, the equality \cite[p.\ 141,
top]{Ziv80} $\hat{I}(Z,U^\ell,X^\ell;V^\ell)=
\hat{I}(Z,X^\ell;V^\ell)$ is questionable because there might be incidental
empirical dependencies between $U^\ell$ and $V^\ell$ given $(Z,X^\ell)$.

Finally, we have concerns regarding the way in which the delay was handled in
\cite{Ziv80}, where the decoder output $v_{t-d}$
was simply renamed $v_t$. It should be kept in mind that
while the data processing theorem applies to $l$--blocks of $\{u_t\}$, $\{x_t\}$,
$\{y_t\}$ and $\{v_{t-d}\}$, the distortion is measured between $u_t$ and $v_t$,
and so, the discrepancy between the $\{v_t\}$ and its delayed version
$\{v_{t-d}\}$ is real
and cannot be handled by simple renaming. Indeed, in \cite{Ziv80}, the
lower bound does not depend on $d$, a fact which is in contrast to the
expectation that the larger is $d$, the better is the performance that can be
achieved.



\begin{thebibliography}{AA}

\bibitem{Atteson99}
K.~Atteson, ``The asymptotic redundancy of Bayes
rules for Markov chains,''
{\it IEEE Trans.\ Inform.\ Theory}, vol.\ 45, 
no.\ 6, pp.\ 2104--2109, September 1999.

\bibitem{CB90}
B.~S.~Clarke and A.~R.~Barron, ``Information--theoretic asymptotics of
Bayes methods,'' {\it IEEE Trans.\ Inform.\ Theory}, vol.\ 36, pp.\ 453-471,
May 1990.

\bibitem{LZ86}
A.~Lempel and J.~Ziv, ``Compression of two--dimensional data,''
{\it IEEE Trans.\ Inform.\ Theory}, vol.\
IT--32, no.\ 1, pp.\ 2--8, January 1986.

\bibitem{Merhav00}
N.~Merhav, ``Universal detection of messages via finite--state channels,''
{\it IEEE Trans.\ Inform.\ Theory},
vol.\ 46, no.\ 6, pp.\ 2242--2246, September 2000.

\bibitem{Merhav13}
N.~Merhav, ``Perfectly secure encryption of individual sequences,''
{\it IEEE Trans.\ Inform.\ Theory}, vol.\ 59, no.\ 3, pp.\ 1302--1310, March
2013.

\bibitem{MW04}
N.~Merhav and M.~J.~Weinberger, ``On universal simulation of information
sources using training data,''
{\it IEEE Trans.\ Inform.\ Theory},
vol.\ 50, no.\ 1, pp.\ 5--20, January 2004.

\bibitem{MZ06}
N.~Merhav and J.~Ziv, ``On the Wyner--Ziv problem for individual sequences,''
{\it IEEE Trans.\ Inform.\ Theory}, vol.\ 52, no.\ 3, pp.\ 867--873, March
2006.

\bibitem{SW73}
D.~Slepian and J.~K.~Wolf, ``Noiseless
coding of correlated information sources,''
{\it IEEE Trans.\ Inform.\ Theory}, vol.\ IT--19, pp.\ 471--480, 1973.

\bibitem{Steinberg09}
Y.~Steinberg, ``Coding and common reconstruction,''
{\it IEEE Trans.\ Inform.\ Theory}, vol.\ 55, no.\ 11, pp.\ 4995--5010,
November 2009.

\bibitem{Ziv78}
J.~Ziv, ``Coding theorems for individual sequences,'' 
{\em IEEE Trans.~Inform.~Theory\/},
vol.~IT--24, no.~4, pp.~405--412, July 1978.

\bibitem{Ziv80}
J.~Ziv, ``Distortion--rate theory for individual sequences,'' 
{\em IEEE Trans.~Inform.~Theory\/},
vol.~IT--26, no.~2, pp.~137--143, March 1980.

\bibitem{Ziv84}
J.~Ziv, ``Fixed--rate encoding of individual sequences with side 
information'',
{\em IEEE Transactions on Information Theory\/},
vol.~IT--30, no.~2, pp.~348--452, March 1984.

\bibitem{ZL78}
J. Ziv and A. Lempel, ``Compression of individual sequences via 
variable-rate coding,''
{\em IEEE Trans.~Inform.~Theory\/},
vol.~IT--24, no.~5, pp.~530--536, September 1978.

\end{thebibliography}
\end{document}